# A free-space coupled, large-active-area superconducting microstrip single-photon detector for photon-counting time-of-flight imaging

Yu-Ze Wang[1], Wei-Jun Zhang[1,2*], Xing-Yu Zhang[1], Guang-Zhao Xu[1], Jia-Min Xiong[1], Zhi-Gang Chen[1,2], Yi-Yu Hong[1,2], Xiao-Yu Liu[1], Pu-Sheng Yuan[1], Ling Wu[1], Zhen Wang[1,2], and Li-Xing You[1,2]

[1]National Key Laboratory of Materials for Integrated Circuits, Shanghai Institute of Microsystem and Information Technology (SIMIT), Chinese Academy of Sciences (CAS), 865 Changning Rd., Shanghai, 200050, China.

[2]Center of Materials Science and Optoelectronics Engineering, University of Chinese Academy of Sciences, Beijing 100049, China.

*E-mail: zhangweijun@mail.sim.ac.cn

**Abstract:** Numerous applications at the photon-starved regime require a free-space coupling single-photon detector with a large active area, low dark count rate (DCR), and superior time resolutions. Here, we developed a superconducting microstrip single-photon detector (SMSPD), with a large active area of 260 μm in diameter, a DCR of ~5 kcps, and a low time jitter of ~171 ps, operated at near-infrared of 1550 nm. As a demonstration, we applied the detector to a single-pixel galvanometer scanning system and successfully reconstructed object information in depth and intensity using a time-correlated photon counting technology.

## 1. Introduction

Single-photon detectors with large active areas are becoming a key technology in many cutting-edge optical applications, spanning from classical to emerging quantum technologies, such as laser ranging and imaging [1-3], biomedical fluorescence imaging [4], and free-space quantum key distribution [5]. Because a large active area significantly helps increase the light-gathering efficiency and reduce the optical path complexity. The most commonly used single photon detectors are the avalanche photodiode (APD) and a photomultiplier tube (PMT), comprised of semiconductor materials and detected using the photoelectric effects. They perform well in visible light detection, but their performance in infrared still needs improvement. For example, APD with a large active area ( >100 μm in diameter) typically has a system detection efficiency (SDE) of 2% at the 1550 nm wavelength, a dark count rate (DCR) of over 100 kcps, and a time jitter of >200 ps [6]. PMT is easy to fabricate with a large active area (>1 mm in diameter), however, its SDE in the near-infrared is <10%, the DCR is in a range of 10k-200k cps, and the timing jitter is >100 ps [7], requiring high operated voltage. In contrast, a superconducting nanowire single-photon detector (SNSPD) has a high SDE (~98%@1550 nm) [8], a low DCR (<1 cps) [9], and low timing jitter (<10 ps) [10] in near-infrared, making it a good candidate for many photon-starved applications. However, generally, an active area of a conventional single-pixel SNSPD is < 20 μm. This is because trade-offs exist between the active area, dark counts, and the timing jitter [11]. Thus, realizing an SNSPD combined with a large active area, a low DCR, and a low timing jitter is challenging.

Generally, a multi-pixel array is an effective method to expand the active area of the SNSPDs [12]. However, arrays comprised of nanowires increase the complexity of fabrication and the difficulty of the readout architectures and have relatively low output pulse amplitudes. Recently, a new class of single-photon detectors with strip widths of several micrometers called superconducting microstrip single-photon detectors (SMSPDs) [13] have become a new option to expand the active area further. In 2018, single-photon sensitivity in SMSPDs was first shown in niobium nitride (NbN) microbridges [14], involving the vortex-assisted mechanism of photon detection. In 2020, the NIST and MIT groups successively used highly disordered amorphous materials (WSi or MoSi) to fabricate SMSPDs with large active areas and saturated detection at 1550 nm [15, 16]. Meanwhile, Vodolazov et al experimentally investigated the timing jitter of an NbN microbridge and theoretically predicted that the lowest time jitter was comparable with that of nanowires [17]. In 2021, Xu et al showed an NbN SMSPD with over 90% SDE using an optical cavity combined with ion-irradiation technology [18]. Later, serval groups have respectively demonstrated that single-pixel SMSPDs can be fabricated with a large active area (up to mm$^2$ scale) using optical lithography (e.g., laser-lithographically written [19] or i-line stepper [15, 20-22]), exhibiting saturated internal detection efficiency (IDE) at near-infrared. Notably, in 2023, through an SMSPD structure and a thermally coupled imager architecture, the NIST and JPL team showed a 0.4-megapixel single-photon camera with an active area of 4 × 2.5 mm$^2$ and a strip width of 1.1 μm [22].

Previously, several SNSPD-based imaging applications have been reported, such as time-of-flight imaging with sub-millimeter depth resolution [10], and polarization resolving and imaging [23, 24]. For wider stripes and larger areas, in 2021, M. Shcherbatenko showed a single-pixel SMSPD camera with an active area of 50×50 μm$^2$ and strip width of 0.41 μm (SDE~18%@1550 nm, 1.7 K), coupling with a multimode fiber and capturing intensity images with single-photon sensitivity [25]. However, there are currently few reports on the applications of SMSPD with large active areas and low-timing jitter, especially for the free-space coupling.

In this study, we designed and fabricated a large active area SMSPD with a diameter of ~ 260 μm and operating at ~ 2.0 K. When adopting the single mode fiber coupling (free-space coupling), the maximum SDE of our SMSPD is ~16% (6%) at 1550 nm, respectively. As a proof-of-principle, we built a free-space, single-pixel scanning galvanometer system, and reconstructed the images using a peak-finding method. The system timing jitter is ~171 ps and the imaging signal-to-background ratio (SBR) is >80. The intensity and depth of the target object are well-reconstructed for the scenes with a 0.5-m stand-away distance, demonstrating the potential of the SMSPD for near-infrared laser ranging and imaging applications at a photon-starved regime.

**2. Design and Fabrication**

The NbN film used in this experiment was deposited on a double-sided polished silicon oxide substrate using a DC reactive magnetron sputtering niobium target in a mixed gas of argon and nitrogen. The thickness of the film was estimated to be about 7 nm according to the deposition time and deposition rate. After deposition, the NbN film was placed in a 300 mm medium current ion implanter (Exceed 2300RD Nissin Co.), and irradiated with 20-keV helium ions under a dose of ion irradiation was $5 \times 10^{16}$ ions/cm$^2$ at room temperature [26]. Electron beam lithography (EBL, with a ZEP resist 1:2) and reactive ion etching (RIE) in $CF_4$ plasma are then used to fabricate the design pattern of microstrips on the irradiated NbN film. Finally, ultraviolet lithography and RIE were used to fabricate coplanar waveguide electrodes.

To enhance the detector's photon absorption and minimize the current crowding effect, we adopted the candelabra-style bends previously reported by Reddy et al [27], as shown in Figure 1. We used COMSOL to simulate the optical absorption efficiency of an NbN-SMSPD, which is embedded in a $SiO_2$/Si semi-optical cavity structure, with a strip width of 2 μm and a high filling factor of 0.8 (see Supplementary Material Figure S2 for more details). Simulation shows that the high-filling factor design has some additional advantages. For example, when the incident light wavelength (*λ*) is 1550 nm, the simulated optical absorption of TE and TM modes are 58.5% and 59.0% respectively, indicating that our detector is polarization-insensitivity. Moreover, the simulated absorption of TE and TM modes also shows a weak dependence on the numerical aperture. Here the width of 2 μm used has considered the trade-off between the kinetic inductance and the IDE of the SMSPD, according to our previous studies [18, 20, 28].

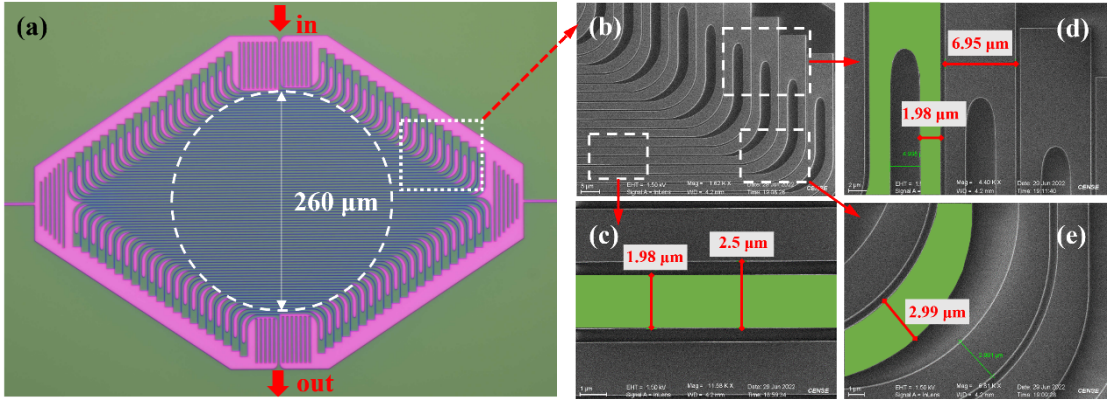

**Figure 1.** (a) Optical microscope image of a candelabra-style SMSPD. The purple area represents the exposed $SiO_2$-coated substrate, and the light-green area represents the NbN film; (b) magnified SEM images of CS-SMSPD near the bends; further zoomed-in SEM images, where the typical sections of the strip are highlighted with light green: (c) the straight part of the microstrip with a high fill factor of 0.8; (d) the 90-degree hairpin turn with a low fill factor of 0.4; (e) the connecting part with a gradient strip width from 2 μm to 3 μm.

Figure 1(a) shows the active area of the fabricated candelabra-style detector (called CS-SMSPD) with a short-axis diameter of ~260 μm and a strip width of 2 μm. Figures 1(b) to 1(e) show the magnified SEM images of the CS-SMSPD: the high filling factor of the straight part of the microstrip is ~0.8, and the gap width is ~0.5 μm (see Figure 1(c)); while the filling factor of the bend area is ~0.4, and the corresponding gap width is ~3 μm (see Figure 1(d)); the straight part and the 90-degree hairpin turn are connected with a gradient width strip, with the widest width of ~3 μm (see Figure 1(e)). For comparisons, we designed a microbridge with the same strip width distributed on the same 2-inch wafer. The width and length of the microbridge are 2 μm and 10 μm respectively, in series with a 6 μm-width NbN meandering strip inductor, providing a kinetic inductance of approximately 1.3 μH. Since the short microbridge has a low probability of fabrication defects, we used the switching current ($I_{sw}$) of microbridges on the same chip as a reference to screen the CS-SMSPDs.

## 3. Device characterization

In this experiment, we used a free-space coupling cryostat based on a pulse tube refrigerator (PT407, Cryomech Co.), with a base temperature of ~ 2 K, a low mechanical vibration (< 2 μm), and a 1-inch optical window [24]. Using an AC coupling readout scheme, the bias current is generated by a DC voltage source (SIM928, SRS Inc.) and a 10-kΩ series resistor, and is input through the DC port of bias-tee (ZX85-12G-S+, Mini Circuit Inc.) to the detector. The response voltage pulse from the detector is transmitted through the capacitor at the RF end of the bias-tee and is amplified by a 50-dB room temperature low-noise amplifier (LNA-650, RF Bay Inc.). The amplified pulse signal is connected to a photon counter (SR400, SRS Inc.) or a high-speed oscilloscope.

Figure 2(a) shows scanning current-voltage (I-V) curves of the adjacent microbridge (blue squares) and CS-SMSPD (red dots) on the same wafer without shunt resistors. $I_{sw}$ of microbridge and CS-SMSPD are 139 μA and 128 μA, respectively. We also characterized the physical parameters of the same microbridge with a physical property measurement system (PPMS, quantum design Co.). Table 1 summarizes the results, showing that the irradiated devices have a critical temperature ($T_c$) of 6.28 K, a high sheet resistance of ~900 Ω/sq, and an estimated $I_{sw}/I_{dep}$ ratio of 0.62 (see Supplementary Material for more details).

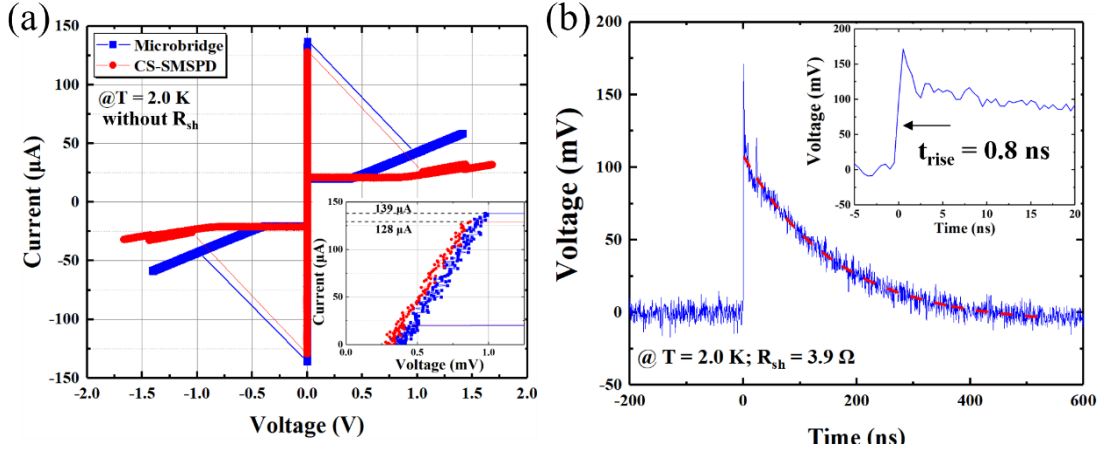

**Figure 2.** (a) Current-voltage (I–V) characteristics of the microbridge and CS-SMSPD measured at 2.0 K. Inset of (a): zoomed-in the I-V curves near the 0-V region; (b) Response pulse waveforms of the CS-SMSPD measured with a shunt resistance of 3.9 Ω and biased at a current of 139 μA (0.9$I_{sw}$)

**Table 1.** Physical parameters of a microbridge fabricated with irradiated NbN thin films at a dose of 5 × 10$^{16}$ ions/cm$^2$. $R_{sq}$ is the square resistance at 20 K, and $D$ is the diffusion coefficient. $I_{dep}$ (0 K) and $I_{dep}$ (2.2 K) are the calculated depairing currents at 0 K and 2.2 K, respectively.

| Sample | $T_c$ (K) | Thickness (nm) | $R_{sq}$ (20 K) (Ω/sq) | $D$ (cm$^2$/s) | $I_{dep}$ (0 K) (μA) | $I_{dep}$ (2.2 K) (μA) | $I_{sw}/I_{dep}$ (2.2 K) |
|---|---|---|---|---|---|---|---|
| Microbridge (irradiated) | 6.28 | 6.5 | 990 | 0.49 | 245.8 | 209.4 | 0.62 |

To prevent detector latching, we connected a shunt resistor directly adjacent to the SMSPD [28, 29]. Then, the packaged SMSPD is mounted on the cold stage of the cryostat and cooled to 2 K. Figure 2 (b) shows the pulse waveform of the SMSPD shunted with a 3.9-Ω resistor and recorded at a bias current ($I_b$) of 139 μA using the oscilloscope. The pulse shows an amplitude of approximately 175 mV with a rising time is about 0.8 ns, and a rest time is about 360 ns (1/$e$ of the falling edge of the pulse). According to the rest time, the kinetic inductance ($L_k$) of the detector is estimated to be approximately 1.40 μH, close to the calculated value of ~1.38 μH using the expression of $L_k = (\hbar R_n)/(1.76kT_c)$ [30]. In our measurements, the readout triggering level is set to 50 mV, greater than the electrical noise floor of <15 mV, ensuring robustness in photon-counting performance.

The SDE of the detector is characterized using a high-precision optical power meter (81624B, Keysight Inc.). A polarization controller (Thorlabs, model FPC561) is employed to adjust the polarization of the input light. The SDE is determined by an expression of SDE = (PCR-DCR)/IPR. Here PCR is the photon count rate when light is incident to the detector. DCR is the dark count rate when a shutter blocks the light. IPR is the input photon rate (~10$^6$ photons/s).

We evaluate the SDE and DCR performance of the detector in single-mode fiber (SMF) coupling and free-space coupling cases, respectively. When using SMF coupling and shunting a 6.8-Ω resistor, the SDE and DCR of the CS-SMSPD are shown in Figures 4(a) and 4(b), separately. When $I_b$ = 146 μA (139 μA), SDE = 15.8% (6.0%) at 1550 nm and 10% (5.8%) at 1064 nm, corresponding to a DCR of ~100 kcps (~5 kcps), separately. To evaluate the IDE of the device, we use the sigmoidal function to fit the SDE-$I_b$ curves. From the fitting results, we estimate that the IDE of the CS-SMSPD is 67%@1550 nm and 76%@1064 nm, respectively. In addition, the saturated values obtained by fitting the sigmoidal function are 23.5% and 13.2%; these two fitting values are in good agreement with the trend of the normalized simulation curve (see Supplementary Material Figure S2(a)). Compared with the IDE (96%@1550 nm) of the microbridge (see Supplementary Material Figure S3), the relatively low IDE of the CS-SMPSD could be due to the increased non-uniformity and defects with the increase in the active area.

We measured the CS-SMSPD in a free-space coupling system at 2.0 K and used a cryolens (~18 mm in diameter) to compress the size of the free-space coupling laser spot to about 65 μm (see Supplementary Material Figure S4 for more details). Here, the shunt resistance was reduced to 3.9 Ω, and the IDE of CS-SMSPD increased to 88%. When $I_b$ = 150 μA, SDE = 5.2% at 1550 nm, with a DCR of 200 kcps. Note that, the efficiency of free space coupling is significantly lower than the result of SMF coupling. We then characterized the optical loss in the coupling process at room temperature using a large-area power meter (S132C, Thorlabs) and found that the optical path loss is ~-0.5 dB, two filters loss is ~-0.5 dB and the cryolens acceptance loss is ~-0.2 dB, resulting in a total coupling loss of ~-1.2 dB.

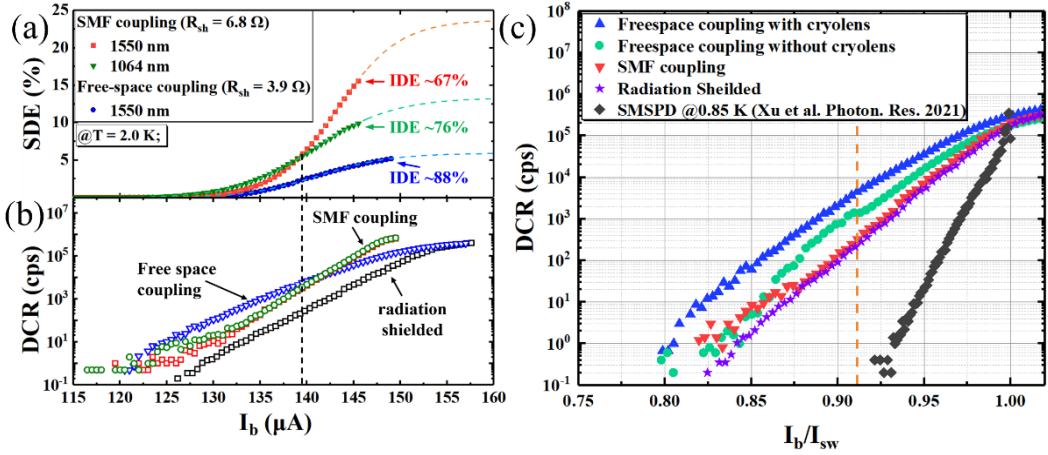

**Figure 3.** (a) SDE as a function of the bias current ($I_b$) for the CS-SMSPD in different coupling cases and varied incident photon wavelengths, operated at a temperature of 2.0 K. Dashed lines are sigmoidal function fits. (b) DCR as a function of $I_b$ for the same CS-SMSPD. (c) DCR performance under different coupling conditions with the normalized bias current. The black dots are adopted from the reference [18], where a 50-in-diameter SMSPD was operated at 0.85 K while shielded from radiation.

Next, we compared the DCR behaviors of the CS-SMSPD. Figure 3(c) shows the DCR measured under different conditions as a function of normalized current. At $I_b/I_{sw}$ = 0.91 (responding to 139 μA), the intrinsic DCR for radiation shielding is approximately 210 cps, while the background DCR for SMF coupling increases to 245 cps. The background dark counts mainly result from blackbody radiations and are linked to the exposed light receiving area's size. Thus, the background DCR of SMSPD increases to 1300 cps under free-space coupling, due to the larger device's exposed active area. When coupled with the cryolens, the DCR rate further rises to 4780 cps, due to further expanding the detector's light-receiving area.

We notice that the intrinsic DCR (~210 cps) is relatively higher than the SNSPD (~0.1 cps) at the same bias current ratio. We suspect the reason is due to the relatively high operating temperature of 2.0 K resulting in relatively high mobility of the thermally activated vortex motion. We expect that lowering the temperature to 0.85 K will significantly improve the dark counting behavior [18], as indicated in Figure 3(c). Moreover, due to the shunt resistance effect, the DCR exhibits a nonlinear region ($I_b/I_{sw}>1$) corresponding to the detector entering an oscillatory state [28], and the detector should be avoided from operating in this region.

Next, we characterized the system time jitter ($\Delta t_{sys}$) of the detector using a time-correlated single photon counting module (TCSPC, SPC-150NX, Becker & Hickl GmbH) [2]. Here, timing jitter is defined as the arrival time deviation of the response pulse from the reference signal, typically represented by the half-width at half-maximum (FWHM) of the cumulative histogram of time deviation, as shown in Figure 4(a). We measured the $\Delta t_{sys}$ as a function of bias current for both CS-SMSPD and microbridge under different coupling conditions. At low bias currents, both devices exhibited significant multipeak histograms, leading to a substantial increase in time jitter. As the bias current gradually increased, the amplitude of the subpeak decreased, resulting in a decrease in the measured $\Delta t_{sys}$ of the main peak, as shown in Figure 4(b).

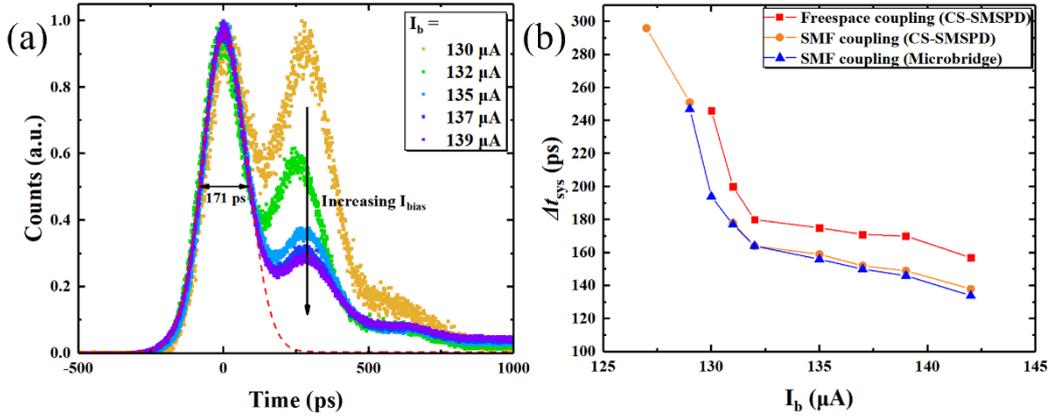

**Figure 4.** (a) Histogram of time-correlated photon counts measured at 1550 nm for different $I_b$. The red dashed line is the Gaussian distribution fit, resulting in an FWHM of 171 ps at $I_b$ = 139 μA. The height of the secondary peak gradually decreases as $I_b$ increases. (b) $\Delta t_{sys}$ of CS-SMSPD and microbridge at varied $I_b$.

Inspired from the analysis method of SNSPD time jitter [17, 31, 32], we express the time jitter of SMSPD as $\Delta t_{sys} = \sqrt{\Delta t_{opt}^2 + \Delta t_{elec}^2 + \Delta t_{ref}^2 + \Delta t_{det}^2}$, where $\Delta t_{opt}$, $\Delta t_{elec}$, $\Delta t_{ref}$, and $\Delta t_{det}$ represent the jitter contributions from optical path, electrical noise, reference signal, and detector's response, respectively. Specifically, $\Delta t_{opt}$ is mainly contributed by the jitter from the dispersion in the optical fiber transmission process (e.g., the multi-mode fiber of 1 meter, jitter <500 fs) and the time difference of the laser source emission (jitter <100 fs). The electrical noise jitter can be estimated by $\Delta t_{ele} = \frac{\sigma_n}{k} \times 2\sqrt{ln2} \approx 110$ ps [31], where $\sigma_n$ is the root mean square value of electrical noise (~10 mV), and $k$ is the slope of pulse edge at the trigger point. Due to the readout scheme with a shunt resistor, the amplitude of the readout pulse signal significantly decreases, resulting in a reduced rising edge slope. $\Delta t_{ref}$ is mainly contributed by the jitter from the TCSPC module (jitter <3.5 ps) and synchronized signal (jitter <0.8 ps).

For SMSPD detectors, $\Delta t_{det} = \sqrt{\Delta t_{geo}^2 + \Delta t_{lat}^2}$, mainly includes the geometric jitter $\Delta t_{geo}$ from differences in signal transmission times due to the length of the strip [32] and the latency jitter $\Delta t_{lat}$ from photon absorption to trigger response [33]. To evaluate the contribution of geometric jitter, in Figure 4(b), we measured the jitter of CS-SMSPD and microbridge under different bias currents using SMF coupling. At the same bias current, we assume that the time jitter contributions, such as electrical noise, are nearly the same for both devices. Therefore, the difference in $\Delta t_{sys}$ between the microbridge and CS-SMSPD is mainly caused by geometric jitter due to different sizes of illuminated areas: (1) in the case of SMF coupling for both devices, the difference of $\Delta t_{sys}$ is approximately 3.1 ps, corresponding to a calculation of $\Delta t_{geo}$ = 30.2 ps and a diameter of the laser spot ~20 μm (see Supplementary Material Inset of Figure S4(a)); (2) comparing $\Delta t_{sys}$ of free-space-coupled CS-SMSPD (blue squares) and SMF-coupled microbridge (red dots), a difference of ~25 ps is obtained at 139 μA, corresponding to an estimated $\Delta t_{geo}$ of ~88 ps, and a diameter of the laser spot of ~65 μm (see Supplementary Material Figure S4(a)); (3) we also evaluate the worst case for the geometric jitter when the light spot illuminates the whole active area of CS-SMSPD, with the jitter value reaching 216 ps, resulting in $\Delta t_{sys}$ increasing to ~264 ps.

Excluding geometric jitter, electrical noise jitter, and other minor contributions, the intrinsic detector's response due to the latency jitter $\Delta t_{lat} = \sqrt{\Delta t_{sys}^2 - \Delta t_{elec}^2 - \Delta t_{ref}^2 - \Delta t_{opt}^2 - \Delta t_{geo}^2}$ provides a jitter of 95.7 ps. To confirm the impact of latency jitter, we measured the system jitter of the microbridge with different incident wavelengths (780 nm, 1064 nm, and 1550 nm, see Supplementary Material Figure S4(b)). When the microbridge was biased at a current of 175 μA and illuminated with photons at 780 nm, saturated IDE was obtained (see Figure S3(a)), the system jitter decreased to 85 ps, and the latency jitter significantly reduced to 37.6 ps. We summarized the contributions of the system timing jitter for the CS-SMSPD and microbridge in Table 2, indicating that the electrical noise jitter, latency jitter, and geometric jitter are the major contributions to our present results.

**Table 2.** Estimated contributions of $\Delta t_{sys}$ for the CS-SMSPD and microbridge. At the same bias current of 139 μA and incident wavelength of 1550 nm, the two devices produced a close latency jitter of 94-96 ps. At a higher bias current of 175 and a shorter incident wavelength of 780 nm, the microbridge demonstrated an improved $\Delta t_{sys}$ of 85 ps and a low latency jitter of 37.6 ps.

| Devices | $I_b$ (μA) | λ (nm) | $\Delta t_{sys}$ (ps) | $\Delta t_{opt}$ (ps) | $\Delta t_{elec}$ (ps) | $\Delta t_{ref}$ (ps) | $\Delta t_{geo}$ (ps) | $\Delta t_{lat}$ (ps) |
|---|---|---|---|---|---|---|---|---|
| CS-SMSPD | 139 | 1550 | 171 | 0.5 (w/ 1m MMF) | 110 | 0.78+3.8 | 88 | 95.7 |
| Microbridge | | | 145 | 0.8 (w/ 4m SMF) | 110 | | 0.01 | 94.3 |
| | 175 | 780 | 85 | 0.8 (w/ 4m SMF) | 76.1 | | | 37.6 |

Considering the impact of dark counts and changes in time jitter, we choose $I_b$ = 139 μA as the operating current for the detector in the following time-of-flight imaging (TOF) demonstration. At this point, the detector's performance exhibits $\Delta t_{sys}$ = 171 ps, corresponding to an SDE of ~2%, and a DCR of ~5 kcps.

## 4. Demonstration of time-of-flight imaging

Figure 5 shows the experimental setup in free-space coupled TOF imaging, consisting of a free-space coupling optical assembly, scanning galvanometer system (GVS112, Thorlabs Inc.), and a cryostat with an optical window (for more details see the Supplementary Material). To simplify, the light is emitted from a synchronized pulse laser and then is adjusted to TE polarization, sequentially passing through a polarizing beam splitter (PBS), two galvanometer mirrors, and reaching the target object. The synchronization signal of the laser as well as the SMSPD output signal is sent to the TCSPC module, where the TOF information is processed through a computer. The inset of Figure 3 demonstrates a focus light spot when using the cryolens, recording through an infrared CCD microscope at room temperature. Not that, at low temperatures, cryostat components' shrinkage and mechanical vibration may cause slight deviations in the detector's position from the room temperature coupling position.

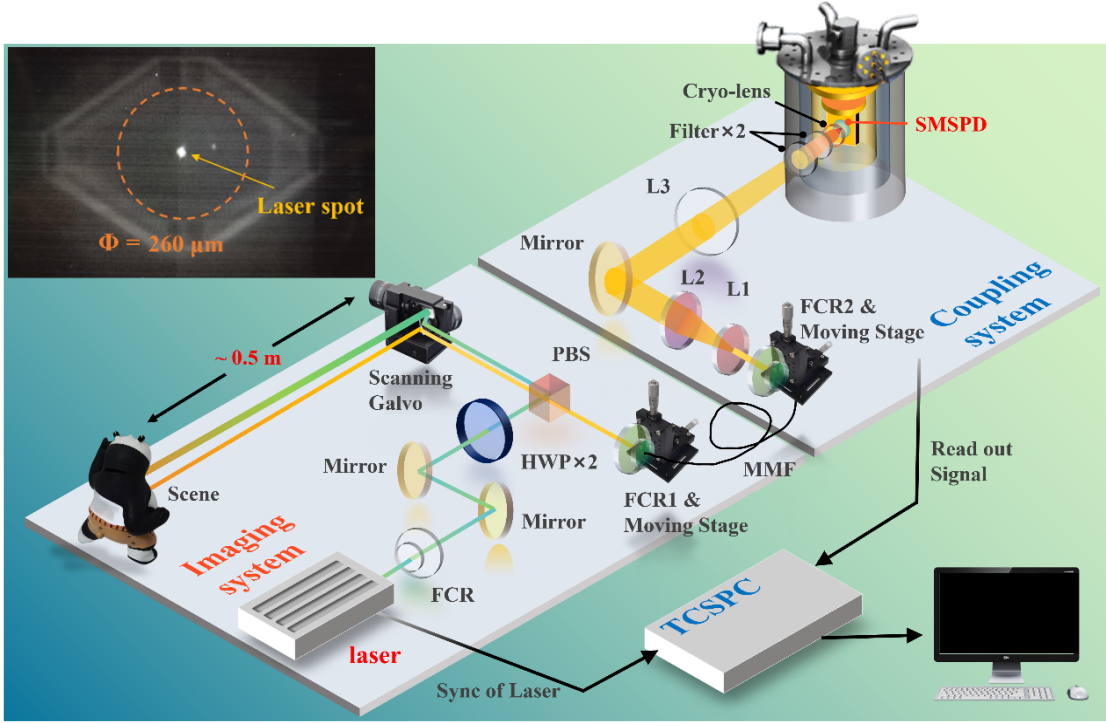

**Figure 5.** Schematic of the time-of-flight imaging setup, mainly including an imaging system and a coupling system. HWP is a wave plate, PBS is a polarization beam splitter, FCR is a fiber collimation package, and MMF is a 1-meter length multimode fiber. Inset: an image of a laser spot coupling to the CS-SMSPD at room temperature using a cryolens, captured by an infrared camera.

Using the scanning system, we can obtain a statistical histogram of the flight time distribution of echo photons. In the experiment, at a distance of 0.5 meters from the scanning galvanometer mirror, the photon count rate (PCR) of the echo signal was 10 kcps -100 kcps, and the DCR corresponded to ~5 kcps. The overall SNR (defined as PCR/DCR) is about 2-20. We adopted an active imaging method in which a pulse light source illuminates the scenes, using the synchronization signal as a reference. By setting an appropriate time window value, the photon signal counts from the background noise counts can be well distinguished, and most of the noise signals are filtered out. For example, when the time window value is 3.3 ns, the background noise counts are lower than 1. Thus, in active imaging, the impact of background dark counts on imaging quality is relatively small.

To extract depth information, a conventional peak-finding method [34] is used to identify the most significant peaks of the echo photon histogram. The resolution of depth imaging is directly related to the timing jitter of the detector. Figures 5(a) and 5(b) show the photos of a large target (toy panda) and a small target (toy leopard). Utilizing the time resolution $\Delta t_{sys}$ of 171 ps and the speed of light $c$ in vacuum, we can estimate the depth resolution in imaging as $\Delta d = 1/2 \times \Delta t \times c \approx 2.5$ cm. Figures 5(c) and 5(d) show the reconstructed depth images. For the larger panda toy, since the depth (in $z$-direction) of the panda toy is ~12 cm, the depth imaging effectively captured the contours and depth variations of the panda toy, such as the actual depth difference between the hand and the belly, which measures approximately 6 cm, corresponding well with the imaging results. While for the leopard toy, the depth imaging only reconstructed the outer contours. The smaller object size coupled with larger jitter results in lower spatial resolution (>2.5 cm), significantly deteriorating the imaging effect.

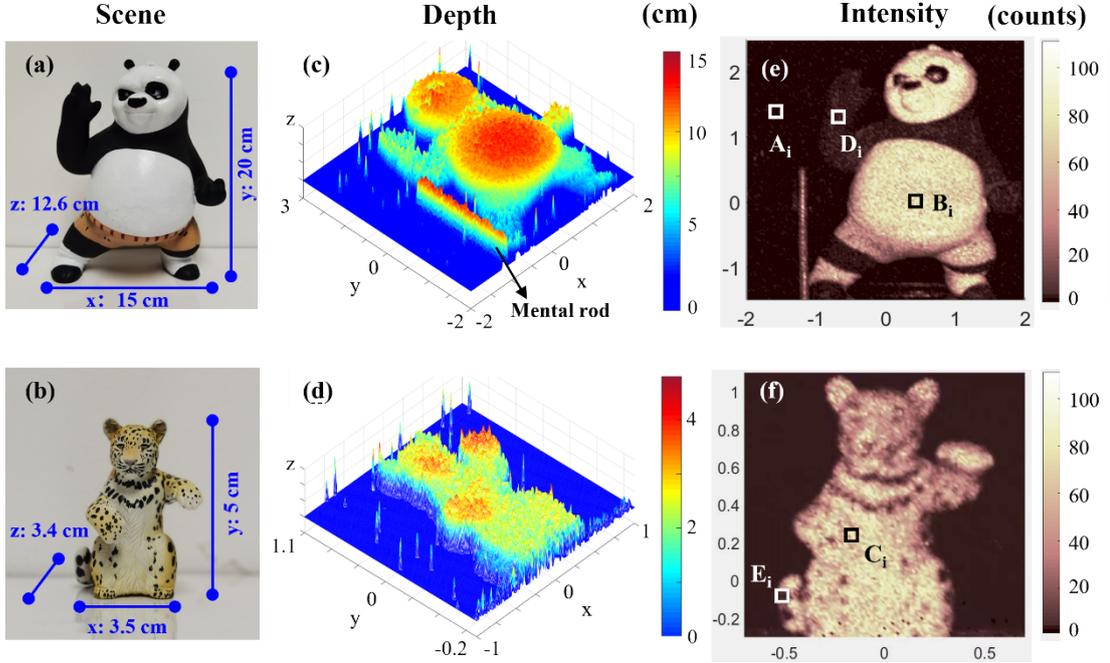

**Figure 6.** Photos of the scenes: (a) a big Panda toy with dimensions of 20 cm ×15 cm ×5 cm, and (b) a small Leopard toy, with dimensions of 0.5 cm × 3.5 cm× 4 cm; reconstructed depth images: Panda (c) and Leopard (d); reconstructed intensity images: (e) Panda and (f) Leopard. The scanning pixel number is 151×151 (101 × 101) for the panda toy (and the leopard toy) respectively, and the dwell time of each pixel is 100 ms.

Figures 5(e) and 5(f) show the reconstructed intensity images, where the contrast of the intensity is directly related to the SNR of the detector. Differences in the color of toys (i.e., reflectivity) will lead to differences in the echo signals of photons, resulting in obvious differences in imaging intensity. For example, in the reconstructed image, the outline of the panda toy and the parts of different colors have been well restored, and the metal rods placed nearby are also clearly displayed (not shown in Figure 5(a)). In the figures, point $A_i$ reflects the background noise level, which is lower than 1 count. The maxima peak height of Points $B_i$ and $C_i$ is about 80 counts. To evaluate our imaging quality, we use the definition of SBR [35], which is the ratio between the maximum peak height and the background level in the histogram of eco photons. It is found that the maximum SBR of the reconstruction is over 80 (i.e., 19 dB) at Points $B_i$ and $C_i$. Although the SBR of points $D_i$ and $E_i$ is about 10 due to their darker color, these two regions can still be distinguished from the background. Notably, in the reconstructed intensity image of the leopard toy, details such as the spots on the leopard's body can be discerned. Because the plane resolution of intensity imaging is related to the plane scanning accuracy of the galvanometer, intensity imaging can well reconstruct the morphological features of the target.

## 5. Discussion

Although we successfully demonstrated photon-counting time-of-flight imaging, there are still some areas in our experiments that need further improvement. First of all, due to the influence of factors such as low IDE, and optical loss in the transmission and coupling, the SDE of our SMSPD is still limited. Secondly, the higher operating temperature results in a relatively higher intrinsic DCR of the detector. Thirdly, a large time jitter due to the detector's latency and electrical noise leads to a relatively large system timing jitter. In the future, we expect to further improve SDE and reduce DCR by some optimizing technologies such as film deposition optimization [36], adopting new geometric designs (such as HCCB structures [37]), and integrating with an optical cavity [18]. Notably, improving the IDE of the detector not only helps to improve the SDE but also helps to suppress the latency jitter of the detector. Using a cryogenic amplifier to reduce electrical noise jitter is also a viable option. We believe that there is still a lot of room for improvement in the performance of the entire free-space system.

## 6. Conclusion

In this work, we developed a large active area SMSPD with a diameter of ~ 260 μm and an operating temperature of ~2.0 K. The maximum SDE of our SMSPD is ~16% (6%) at 1550 nm, corresponding to single-mode fiber coupling (free-space coupling), respectively. With a DCR of 5 kcps, the SDE of the free-space coupled detector is 2% and the system timing jitter is about 171 ps. We analyzed the influencing factors of SMSPD's time jitter, where the main contributions are the electrical noise jitter of ~110 ps, the latency jitter of ~96 ps, and the geometric jitter of ~88 ps (coupled with a small light spot using cryolens). We utilized this detector in a free-space single-pixel scanning system, achieving imaging reconstruction in depth and intensity for scenes at a 0.5 m stand-off distance at the low-light level. The depth resolution is about 2.5 cm, and the maximum SBR for intensity imaging is about 80. Our research demonstrates the potential application prospects of SMSPD with large active area and low timing jitter in laser ranging, imaging, and related fields.

**Acknowledgments**

This work is supported by the National Natural Science Foundation of China (NSFC, Grant No. 62371443 and 61971409), the Science and Technology Commission of Shanghai Municipality (Grant No. 18511110202 and No. 2019SHZDZX01), and the Shanghai Sailing Program (Grants No. 21YF1455500). W. -J. Zhang is supported by the Youth Innovation Promotion Association (No. 2019238), Chinese Academy of Sciences.

**Author contributions:** W.-J.Z. and Y.-Z.W. conceived and designed the experiments. Y.-Z.W., X.-Y.Z., and G.-Z.X. carried out the experiments. Y.-Z.W. and W.-J.Z. analyzed the data and wrote the manuscript with the input of all authors. All authors discussed the results and reviewed the manuscript.

**Competing interests:** The authors declare no competing interests.

**Data and materials availability:** Data underlying the results presented in this paper are not publicly available at this time but may be obtained from the authors upon reasonable request.